\documentclass[twocolumn,showpacs,preprintnumbers,amsmath,amssymb]{revtex4}

\usepackage{graphicx}
\usepackage{dcolumn}
\usepackage{bm}
\usepackage{epstopdf}

\begin{document}

\preprint{}

\title{Femtosecond relativistic electron beam with reduced timing jitter from THz-driven beam compression}

\author{Lingrong Zhao$^{1,2}$, Heng Tang$^{1,2}$, Chao Lu$^{1,2}$, Tao Jiang$^{1,2}$, Pengfei Zhu$^{1,2}$, Long Hu$^{3}$, Wei Song$^{3}$, Huida Wang$^{3}$, Jiaqi Qiu$^{4}$, Chunguang Jing$^{5}$, Sergey Antipov$^{5}$, Dao Xiang$^{1,2,6*}$ and Jie Zhang$^{1,2*}$}
\affiliation{%
$^1$ Key Laboratory for Laser Plasmas (Ministry of Education), School of Physics and Astronomy, Shanghai Jiao Tong University, Shanghai 200240, China \\
$^2$ Collaborative Innovation Center of IFSA (CICIFSA), Shanghai Jiao Tong University, Shanghai 200240, China \\
$^3$ Science and Technology on High Power Microwave Laboratory, Northwest Institute of Nuclear Technology, Xi'an, Shanxi 710024, China\\
$^4$ Nuctech Company Limited, Beijing, 100084, China\\
$^5$ Euclid Techlabs LLC, Bolingbrook, Illinois 60440, USA \\
$^6$ Tsung-Dao Lee Institute, Shanghai 200240, China\\
}
\date{\today}

\begin{abstract}
We propose and demonstrate a novel method to reduce the pulse width and timing jitter of a relativistic electron beam through THz-driven beam compression. In this method the longitudinal phase space of a relativistic electron beam is manipulated by a linearly polarized THz pulse in a dielectric tube such that the bunch tail has a higher velocity than the bunch head, which allows simultaneous reduction of both pulse width and timing jitter after passing through a drift. In this experiment, the beam is compressed by more than a factor of four from 130 fs to 28 fs with the arrival time jitter also reduced from 97 fs to 36 fs, opening up new opportunities in using pulsed electron beams for studies of ultrafast dynamics. This technique extends the well known rf buncher to the THz frequency and may have a strong impact in accelerator and ultrafast science facilities that require femtosecond electron beams with tight synchronization to external lasers. 
\end{abstract}

\maketitle

Ultrashort electron beams with small timing jitter with respect to external lasers are of fundamental interest in accelerator and ultrafast science communities. For instance, such beams are essential for laser and THz-driven accelerators (\cite{DLARMP, THzNC, THzIFEL, THzNP}) where the beam energy spread and beam energy stability largely depend on the electron bunch length and injection timing jitter, respectively. For MeV ultrafast electron diffraction (UED \cite{UED3, UCLA, THU, OSAKA, SJTU, BNL, SLAC, DESY}) where ultrashort electron beams with a few MeV energy are used to probe the atomic structure changes following the excitation of a pump laser, the temporal resolution is primarily limited by the electron bunch length and timing jitter. Similar limitations exist for x-ray free-electron lasers (FEL \cite{LCLS, SACLA, PAL}) too, since the properties of the x-ray pulses depend primarily on that of the electron beams. Therefore, one of the long-standing goals in accelerator and ultrafast science communities is to generate ultrashort electron beams with small timing jitter. 

    \begin{figure*}[t]
    \includegraphics[width = 0.85\textwidth]{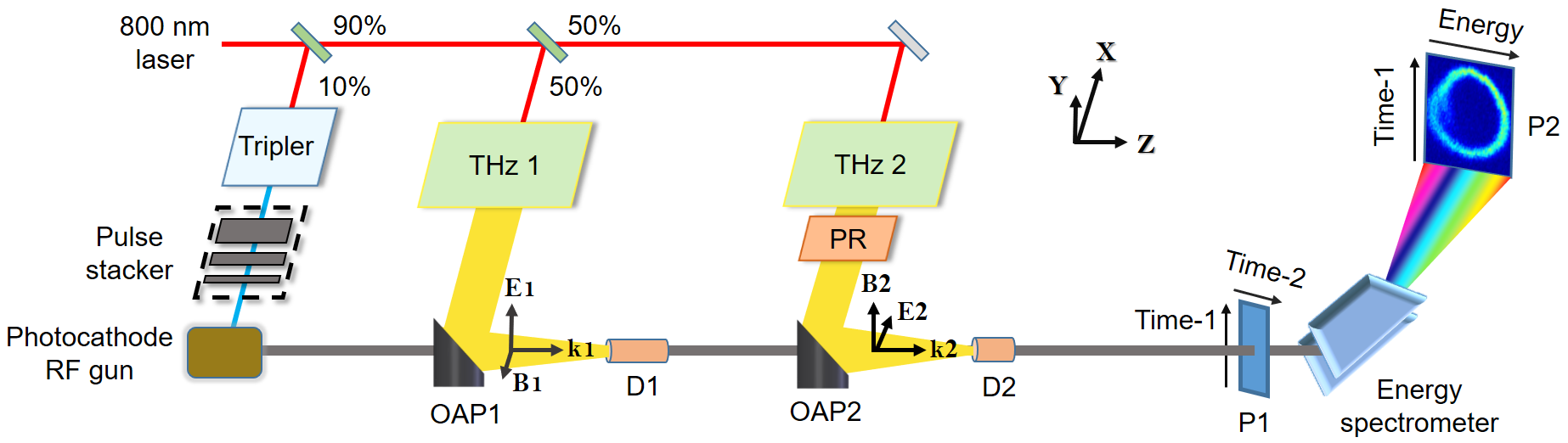}
            \caption{THz-driven relativistic electron beam compression experiment setup. The 3.0 MeV electron beam is compressed by the first THz pulse with vertical polarization and measured by the second THz pulse with horizontal polarization. The THz pulses are focused to the dielectric tubes (D1 and D2) with off-axis parabolic mirrors (OAP) in colinear configurations. 
    \label{Fig.1}}
    \end{figure*}

Photocathode rf gun is the leading option for producing high brightness ultrashort electron beam for FEL and MeV UED (see, e.g. \cite{KoreaUED, LCLSinjector}). Due to space charge effect, the electron beam pulse width is broadened and therefore bunch compression is typically needed to reduce the pulse width. Bunch compression requires first a mechanism to imprint energy chirp (correlation between an electron's energy and its longitudinal position) and then sending the beam through a dispersive element such that the longitudinal displacement of the electrons is changed in a controlled way for reducing the pulse width. For MeV beam, this is typically achieved by first sending the beam through a rf buncher cavity at zero-crossing phase where the bunch head ($t<0$) is decelerated while the bunch tail ($t>0$) is accelerated. This imprints a negative chirp $h=d\delta/cdt<0$ in the beam longitudinal phase space, where $\delta$ is the relative energy difference of an electron with respect to the reference electron and $c$ is the speed of light. Then the electron beam is sent through a drift after which the electrons at the bunch tail catch up with those at the bunch head, leading to compression in pulse width. Full compression is achieved when $hR_{56}$=-1, where $R_{56}\approx L/\gamma^2$ is the momentum compaction of the drift with length $L$ and $\gamma$ is the Lorentz factor of the electron beam. 

Recently, sub-10 (rms) fs beams have been produced with this rf buncher technique \cite{CUCLA, PRX}. However, the rf phase jitter results in increased beam timing jitter after compression. While THz pulse based time-stamping techniques have been developed to measure and correct the electron beam arrival time jitter \cite{PRX, SLACstreaking} for UED, the detector response time limited this shot-to-shot correction technique to low repetition rate. It is highly desired (in particular for those experiments that require long data acquisition time (see, e.g. \cite{SLACgasphaseCF3I})) if the beam can be compressed without increasing the jitter such that time-stamping technique is not needed. Many efforts have been devoted towards this goal in the past few years. For instance, it has been shown that replacing the rf buncher with a laser-driven THz buncher can be used to compress keV beams while simultaneously keeping the timing jitter below 10 fs \cite{THzNP, THzbuncher1, THzbuncher2}, making full use of the fact that the THz pulse is tightly synchronized with the laser. However, with the THz pulse propagating perpendicularly to the electron beam path, a very strong THz source is required for compressing a MeV beam because the interaction length is rather limited.  

In this Letter, we demonstrate a novel method to compress a relativistic electron beam using a THz pulse with moderate strength. In this technique the THz pulse co-propagates with the electron beam in a dielectric tube and the interaction length is thus greatly increased. The schematic layout of the experiment is shown in Fig.~1. An 800 nm laser is split into three pulses with one pulse (1 mJ) used for producing electron beam in a photocathode rf gun and the other two pulses (4 mJ each) for producing THz radiation (about 1.5 $\mu$J each) through optical rectification in LiNbO$_3$ crystal \cite{TPFP}. The first THz pulse with vertical polarization is injected into a dielectric tube where it interacts with the electron beam and imprints an energy chirp in beam longitudinal phase space. The beam is then compressed after passing through a 1.4 m drift where the bunch length and arrival time jitter are measured with a second THz pulse that deflects the beam in horizontal direction. The electron beam can be measured either on a screen (P1) before the energy spectrometer or after it (P2). 

The field pattern in a conventional rf buncher is typically TM01 mode for which the longitudinal field has a very weak dependence on transverse position, ideal for producing energy chirp without increasing beam uncorrelated energy spread. However, efficient excitation of TM01 mode typically requires an input pulse with radial polarization and it is difficult to excite such mode with a linearly polarized THz pulse \cite{MC}. In our experiment, a vertically polarized THz pulse is directly injected into a dielectric tube (D1) and thus only HEM11 mode is excited. Such a mode has recently been used to deflect electron beam for measuring bunch length and timing jitter \cite{THzO}. It should be noted that HEM11 mode has longitudinal electric field which varies linearly with transverse offset. This effect may be understood with Panofsky-Wenzel theorem \cite{PW} which connects the time-dependent angular kick with offset-dependent energy kick. So energy chirp may be produced with this longitudinal electric field when the electron beam passes through the dielectric tube off-axis, allowing THz-driven bunch compression with HEM11 mode.

In our experiment D1 is a $L_D$=15 mm long cylindrical quartz tube with inner diameter of 910 $\mu$m and outer diameter of 970 $\mu$m. The outer surface of the tube is gold coated and analysis shows that such structure can support HEM11 mode with frequency at about 0.66 THz. To clearly show the effect of the longitudinal electric field, we used a set of BBO crystals to stack the UV laser pulse for producing a long electron beam with pulse width comparable to the wavelength of the HEM11 mode. Representative beam distributions when it passes through D1 with various offsets are measured at screen P2 and shown in Fig.~2a. Because the beam is deflected in vertical direction and bent in horizontal direction, the horizontal axis on screen P2 becomes the energy axis and vertical axis becomes the time axis. From Fig.~2a, one can see that when the electron beam enters the tube on-axis ($y_0=0$), the beam is only deflected by the THz pulse and the distribution takes a stripe shape. With the electron pulse width comparable to the wavelength of HEM11 mode, the sinusoidal deflection leads to a double-horn distribution. When the beam passes through the tube off-axis, the beam receives time-dependent energy kick from the longitudinal electric field and thus takes a ring shape. Specifically, for $y_0<0$, region A represents the electrons that experience acceleration phase; electrons in region C experience deceleration phase; electrons in regions B and D are at the zero-crossing phases and thus the centroid energy is not changed. As the offset is gradually changed from $-200~\mu$m to $150~\mu$m, the distance between regions A and C first reduces and then increases again after reaching a minimum at $y_0=0$. Furthermore, after the sign of the offset is reversed ($y_0>0$), electrons in region A are now decelerated while those in region C are accelerated, consistent with the fact that the longitudinal field scales linearly with transverse offset.

    \begin{figure}[t]
    \includegraphics[width = 0.49\textwidth]{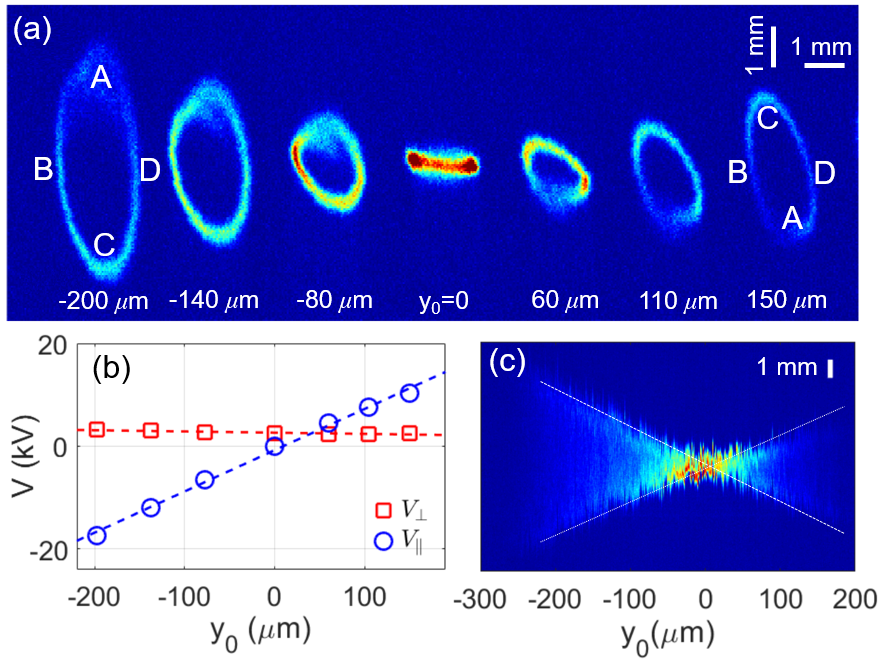}
            \caption{Beam distributions at screen P2 (a) and longitudinal and transverse energy kicks (b) for various offsets in D1. Note, for the purpose of illustration, the horizontal axis in (a) represents time and vertical axis represents energy. The superimposed beam distribution on P2 for a fine scan of offsets in D1 is shown in (c).   
    \label{Fig.2}}
    \end{figure}

The maximal energy change of the electrons can be found using the distance between A and C in Fig.~2a. Similarly, the maximal transverse energy kick can be found using the distance between B and D. After converting the distance into voltage using the known dispersion, the longitudinal energy kick from longitudinal electric field ($V_{\parallel}$) and transverse energy kick ($V_{\perp}$) from transverse electromagnetic field are shown in Fig.~2b. The transverse kick is independent of the transverse offset and the longitudinal kick scales linearly with the offset. In a separate experiment, the beam distribution at P2 is measured with a fine scan of $y_0$ and the distributions are then superimposed and shown in Fig.~2c where a cone shape (guided with the dotted line) is clearly seen that shows the linear dependence of energy change on transverse offset.  

A closer look at the rings in Fig.~2a indicates that the transverse kick and longitudinal kick have $\pi/2$ phase difference. For instance, electrons in regions A and C experience on-crest phase for acceleration but zero-crossing phase for deflection. To confirm this, the pulse stacker is removed and now a beam with about 130 fs pulse width is produced. The timing of the THz beam is varied and the beam centroid change as well as energy change are measured simultaneously and the results are shown in Fig.~3 which confirms the $\pi/2$ phase difference in acceleration and deflection. This $\pi/2$ phase difference results in a net shift of the beam centroid divergence during compression and thus can be straightforwardly corrected with a steering magnet. With the group velocity of the THz pulse being about $v_g\approx0.84$c in D1, the interaction window within which the electrons can catch up and interact with the THz pulse in D1 is about $(c/v_g-1)L_D/c\approx10~$ps. This is the main reason that multiple oscillations are observed in Fig.~3 even when a single-cycle THz pulse is used.

    \begin{figure}[b]
    \includegraphics[width = 0.49\textwidth]{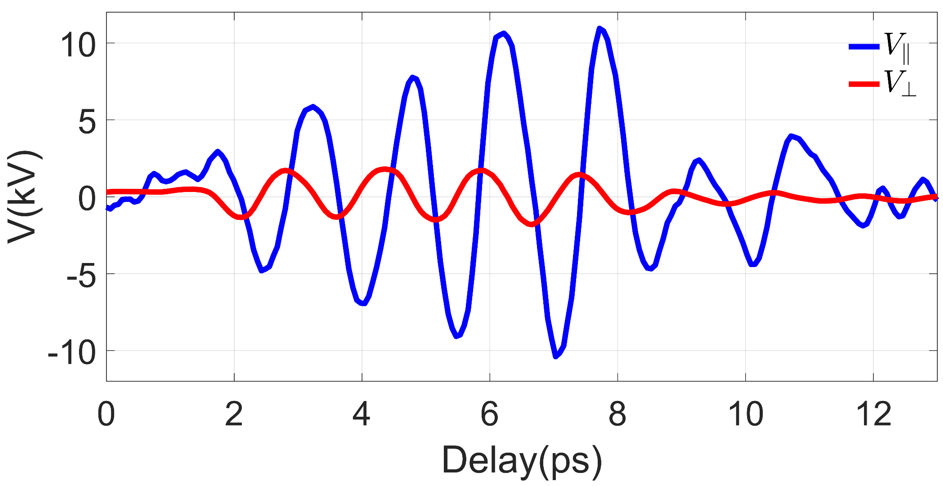}
            \caption{Longitudinal and transverse energy kicks as a function of time delay between the first THz pulse and electron beam.
    \label{Fig.3}}
    \end{figure}

To measure the time information of the compressed bunch, a second THz pulse is used to deflect the beam for converting time information into spatial distribution. In our first attempt, the second THz pulse is also vertically polarized as the crystals and gratings have the same configurations for the two THz sources. However, it is noted that such a configuration may introduce ambiguity in data interpretation. For instance, when the electron beam passes through both tubes on-axis, but with $\pi$ difference in phase, then the deflection from the first THz pulse is canceled by the second THz pulse. This produces a transverse beam size comparable to that with full compression while apparently no compression occurs here. To allow unambiguous determination of the bunch compression effect, in our experiment a polarization rotation (PR) element consisting of a pair of wire grid polarizers and roof mirrors are used to manipulate the THz polarization. As shown in \cite{THzO}, by changing the path length of the two roof mirrors, a vertically polarized THz pulse can be converted into a horizontally polarized pulse. 

Then we sent the electron beam through D1 at an offset of 200 $\mu$m, and used a steering magnet to center the beam in the second tube (D2). The beam distributions on screen P1 measured with the two THz pulses on for various time delay between the electron beam and first THz pulse are shown in Fig.~4. In this measurement the timing of the second THz pulse is adjusted such that the beam always rides at the zero-crossing phase of the deflection field. Fig.~4a-d corresponds to the cases when the electron beam is at regions A, B, C, and D in Fig.~2a, respectively. As shown in Fig.~4a and Fig.~4c, the beam is at on-crest phases for acceleration and at the zero-crossing phases for deflection in D1, so the beam is deflected vertically by the first THz pulse and deflected horizontally by the second THz pulse, leading to correlation in horizontal and vertical distribution. The beam is not compressed at these time delays. It should be noted that the streaked beam size is larger than the aperture of D2, so only part of the beam is measured on screen P1. The beam is at one of the zero-crossing phases for acceleration in Fig.~4b. However, at such phase the bunch head is accelerated while the bunch tail is decelerated. As a result, a positive energy chirp is imprinted in the beam phase space, leading to bunch lengthening by roughly a factor of 2. The bunch duration in this case is larger than the dynamic range of the measurement (roughly one quarter of the deflection wavelength), and thus a quasi-flattop distribution is seen for the deflected beam. 

Fig.~4d shows the distribution when the beam is at the right zero-crossing phase for bunch compression. For this case the beam is shorter than the dynamic range of the measurement and thus the bunch length can be accurately determined. In this measurement the ramping rate of the deflection from the second THz pulse is found to be about 4 $\mu$m/fs. With the transverse beam size and beam centroid fluctuation at screen P1 measured to be about 120 $\mu$m and 16 $\mu$m with the THz off, the resolution of beam temporal profile measurement and the accuracy of beam arrival time measurement are determined to be about 30 fs and 4 fs, respectively. With the strength of the first THz pulse varied to provide the optimal energy chirp, full compression is achieved and the raw beam pulse width in Fig.~4d after converting the beam size to time is measured to be about 41 fs (rms). Subtracting the resolution in quadrature yields a true bunch length of about 28 fs (rms).  

    \begin{figure}[t]
    \includegraphics[width = 0.49\textwidth]{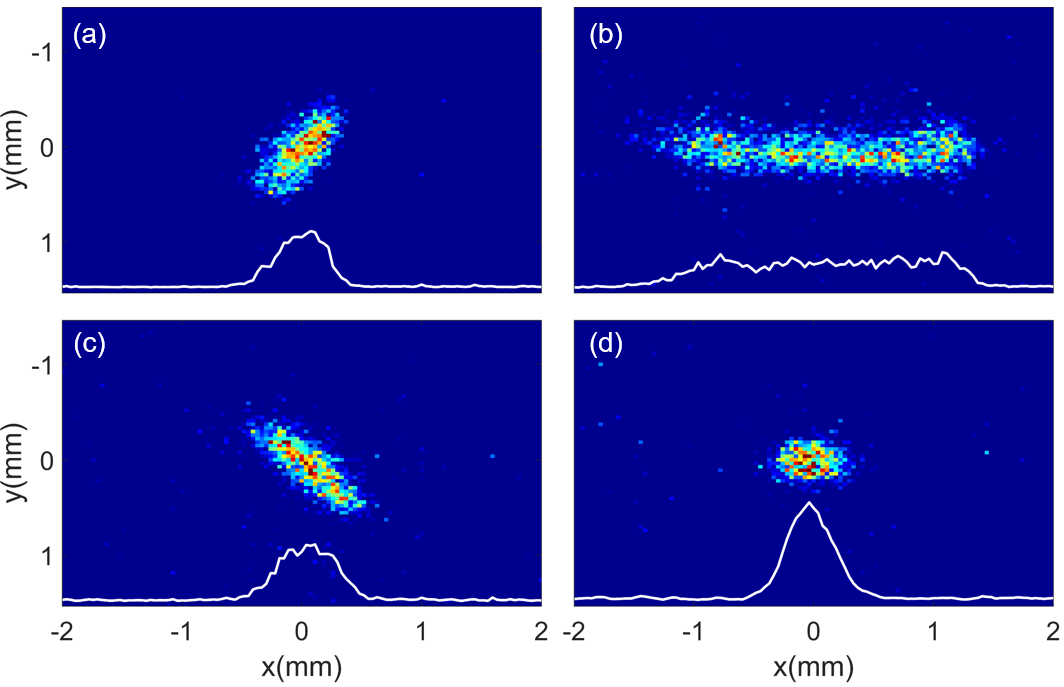}
            \caption{Beam distributions at screen P1 measured with both THz pulses on for various time delays between the first THz pulse and the electron beam. 
    \label{Fig.4}}
    \end{figure} 

Under full compression condition, 50 consecutive measurements of the raw beam profile at P1 (with horizontal axis converted into time) with THz buncher off and on are shown in Fig.~5a and Fig.~5d, respectively. The effect of longitudinal compression can be clearly seen. The average bunch length (calculated by subtracting the contribution from intrinsic beam size) before and after THz compression is about 130 fs and 28 fs, respectively. The fluctuation of the beam centroid (black dots in Fig.~5a and Fig.~5d) which represents the time jitter with respect to the second THz pulse is also greatly reduced from about 97 fs to about 36 fs after compression.  

    \begin{figure}[b]
    \includegraphics[width = 0.49\textwidth]{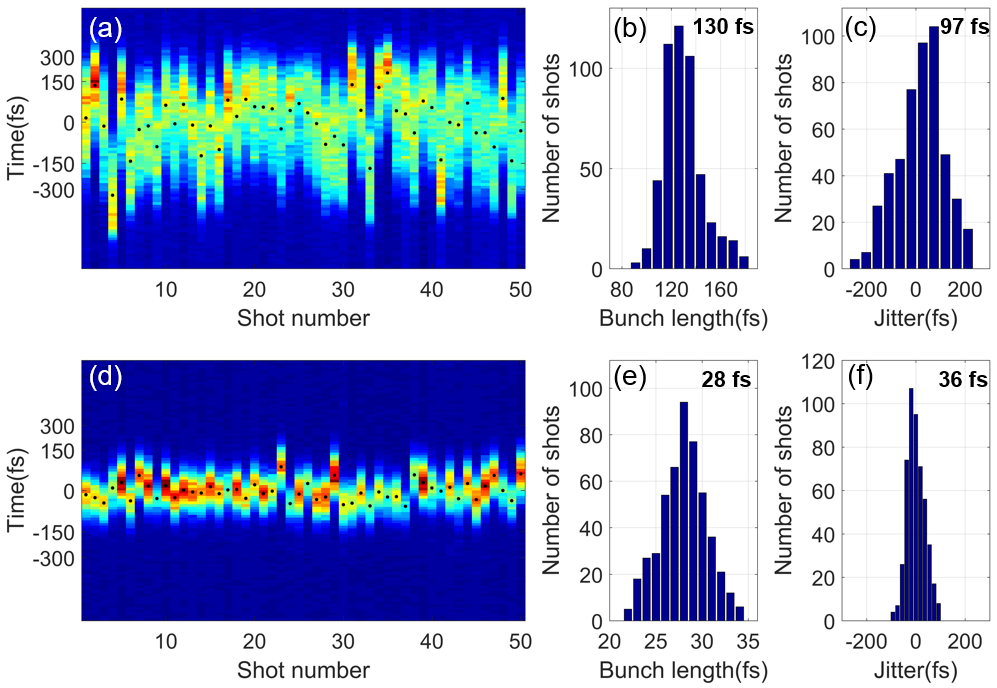}
            \caption{50 consecutive single-shot measurements of the beam temporal profile without (a) and with (d) the THz buncher, and the rms pulse width statistics collected over 500 consecutive shots without (b) and with (e) the THz buncher. The beam centroid for each individual shot (black dots) is used to determine the time jitter without (c) and with (f) the THz buncher. 
    \label{Fig.5}}
    \end{figure} 

It should be noted that while the initial jitter before the THz buncher is greatly compressed, the energy stability of the beam still limits the residual timing jitter of the THz buncher scheme to $\Delta t=R_{56}\delta E/E$, where $\delta E/E$ is the relative energy stability of the beam. In this experiment, the beam energy stability is measured to be about $2.5\times10^{-4}$, which together with 3.9 cm momentum compaction from D1 to D2 leads to a residual timing jitter of about 32 fs, in good agreement with the experimental result. For keV UED with high energy stability, such timing jitter is negligible. With full compression, the residual bunch length is limited by the uncorrelated beam energy spread ($\sigma_{\delta}$) to $\sigma_{\delta}R_{56}/E$. In our experiment, the beam uncorrelated energy spread grows in the THz buncher because of the linear dependence of the longitudinal electric field and the finite transverse beam size (about 20 $\mu$m) in D1. The uncorrelated energy spread growth is estimated to be about 0.55 keV which results in a residual bunch length of about 24 fs, consistent with the measured value. The minimal bunch length may be further reduced by focusing the beam to a smaller size. It is also possible to use an additional dielectric tube to cancel the energy spread growth. For instance, after the THz buncher, the beam may be sent through a second tube on-axis. By using a THz pulse with $\pi$ phase difference, the energy spread growth may be effectively canceled with the chirp unchanged. The residual timing jitter may be reduced with improved energy stability, or using a THz pulse with stronger strength \cite{GV} for lowering the required momentum compaction. Such steps should be able to push both the electron bunch length and arrival time jitter to sub-10 fs regime, opening up new opportunities in ultrafast science and advanced acceleration applications.  

In conclusion, we have demonstrated a novel method to manipulate relativistic beam phase space for longitudinal compression. By using the longitudinal field in HEM11 mode for imprinting the energy chirp and with the THz pulse co-propagating with the electron beam, the mode is easily excited with a linearly polarized THz pulse and the required THz pulse energy for producing sufficient energy chirp is greatly reduced. In our experiment we have demonstrated significant reduction in both bunch length and arrival time jitter, which may allow one to significantly enhance the temporal resolution of UED. The demonstrated colinear interaction scheme is also of interest for THz-driven beam acceleration that holds potential for downsizing accelerator-based large scientific facilities such as FELs and colliders. We expect this THz-driven beam manipulation method to have wide applications in many areas of researches. 

This work was supported by the Major State Basic Research Development Program of China (Grants No. 2015CB859700) and by the National Natural Science Foundation of China (Grants No. 11327902, 11504232 and 11721091). One of the authors (DX) would like to thank the support of grant from the office of Science and Technology, Shanghai Municipal Government (No. 16DZ2260200 and 18JC1410700).\\
* dxiang@sjtu.edu.cn \\
* jzhang1@sjtu.edu.cn

\pagebreak

\end{document}